\documentclass[twocolumn,showpacs,preprintnumbers,amsmath,amssymb,prb,aps]{revtex4}

\usepackage{epsfig}
\usepackage{graphicx}
\usepackage{dcolumn}
\usepackage{bm}

\begin{document}

\title{Electronic and structural transition in La$_{0.2}$Sr$_{0.8}$MnO$_3$}

\author{R. Bindu and Kalobaran Maiti}
\affiliation{Department of Condensed Matter Physics and Materials
Science, Tata Institute of Fundamental Research, Homi Bhabha Road,
Colaba, Mumbai - 400 005, INDIA.}

\author{R. Rawat}
\affiliation{UGC-DAE Consortium for Scientific Research, University
Campus, Khandwa Road, Indore 452 017, India.}

\author{S. Khalid}
\affiliation{National Synchrotron Light Source, Brookhaven National
Laboratory, Upton, New York - 11973}

\date{\today}

\begin{abstract}

We investigate the interplay of the electronic and structural
transition in La$_{0.2}$Sr$_{0.8}$MnO$_{3}$. The transport and
specific heat measurements exhibit unusual evolutions and signature
of a first order phase transition around 265 K. Mn K-edge extended
$x$-ray absorption fine structure results reveal distortion in the
MnO$_6$ octahedra even in the cubic phase and a remarkable evolution
of the distortion across the phase transition. These results
manifest the importance of fluctuations in Mn 3$d$ orbital occupancy
and disorder in their electronic properties, which may help in
understanding the orbital and spin ordering proposed in these
systems.

\end{abstract}

\pacs{61.50.Ks, 75.30.Kz, 75.47.Lx, 61.05cj}

\maketitle


Discovery of colossal magnetoresistance (CMR) in manganites have
opened up a new field of research in solid state physics due to
their interesting fundamental issues and significant technological
applications such as high density storage media, field sensors, {\it
etc}. These systems exhibit plethora of exotic properties such as
charge, orbital and spin ordering, electronic phase separation, {\it
etc.} in addition to the CMR effect. It is now realized that the
intricate competition between the charge, spin, orbital, and lattice
degrees of freedom in these mixed valent manganites are important
for these exotic properties.\cite{coey,salamon,dagotto}

Among all the manganites, La$_{1-x}$Sr$_{x}$MnO$_{3}$ ($x <$ 0.5) is
studied the most due to its proximity to the CMR effect. With the
increase in $x$, the tolerance factor becomes close to unity and the
Mn-O-Mn angle also becomes close to 180$^o$. In this situation, Mn
$e_{g}$ orbitals would be almost degenerate\cite{brink,ryo} and is
ideal to study the electronic charge distribution in Mn 3$d$ levels
across the electronic phase transitions observed in these systems.

In this paper, we report our results on the electronic and
structural evolution associated to MnO$_6$ octahedra in
La$_{0.2}$Sr$_{0.8}$MnO$_3$ as a function of temperature. This
compound exhibits a transition from paramagnetic to C-type
antiferromagnetic (C-AFM) structure around 265 K.\cite{ch} Reports
on structural transition are controversial suggesting a cubic to
tetragonal\cite{ch} or no structural transition\cite{hem} with
temperature. Subsequent theoretical studies\cite{brink,ryo} predict
importance of orbital degeneracy in this system. The orbital
degeneracy of Mn 3$d$ states strongly depends on the local crystal
structure of the MnO$_6$ octahedra, which can be probed efficiently
by extended $x$-ray absorption fine structure (EXAFS) technique.
Thus, we have employed Mn K-edge EXAFS around the region of
electronic phase transition to investigate the evolution of the
local structural parameters of the MnO$_6$ octahedra. Our results
reveal that the MnO$_{6}$ octahedra are distorted even in the cubic
phase. The results at lower temperatures manifest phenomenal
evolution of the Mn-O bond lengths across the phase transition.


The sample was prepared by conventional solid state route
\cite{EPJB} and characterized by powder $x$-ray diffraction
technique. Sample was found to be in single phase with no signature
of impurity feature. The analysis of the diffraction peaks at room
temperature indicates cubic structure with space group {\it
Pm$\overline{3}$m}. Iodometric redox titration confirmed the sample
to be stoichiometric. The resistivity and heat capacity measurements
were performed using standard four probe method and home made semi
adiabatic heat pulse calorimeter, respectively. Mn K-edge EXAFS
studies were carried out at X-18B beamline at the National
Synchrotron Light Source, Brookhaven National Laboratory with an
eneregy resolution of 0.8 eV at Mn K-edge.

EXAFS analysis was carried out using UWXAS software\cite{stern}
following the standard procedure.\cite{xray} The Fourier transform
(FT) to the $R$-space was carried out in the $k$-range 3 - 13
\AA$^{-1}$ by Fourier transforming $k^{2}\chi(k)$ ($\chi(k)$ is the
EXAFS oscillations) by a Hanning window function. The back
scattering amplitude and phases were calculated using FEFF6.01a
code\cite{zabinsky} for LaMnO$_{3}$ and the same were used for all
the temperatures. The first coordination shell fit was done in the
Fourier filtered $k$-space in the range 0.86 - 1.87 \AA. The overall
many body reduction factor, $S_0^2$ was fixed to 0.82. As seen
earlier\cite{JPCM}, we find that the Mn K-edge EXAFS data at room
temperature are best described by 2 short Mn-O bond lengths, $R_1$
and 4 long Mn-O bond lengths, $R_2$. In order to avoid uncertainty
in the fitting procedure, we have considered average bond distances
for eight Mn-Sr/La bonds and six Mn-Mn bonds in higher coordination
shells (0.86 - 3.87 \AA).


The temperature dependent resistivity, $\rho$ and heat capacity,
$C_p$ are plotted in Fig. 1. $\rho$ in Fig. 1(a) exhibits insulating
temperature dependence along with a sharp increase at about 265 K
($T_1$) and a change of slope at about 254 K. In addition, a
distinct hysteresis is observed around $T_1$ and 224 K ($T_2$) for
the cooling and warming cycles. This is demonstrated in Fig. 1(b),
where we show the difference in $\rho$ between cooling ($\rho_c$)
and warming ($\rho_w$) cycles normalized by $\rho_c$. It is evident
that $(\rho_c - \rho_w)/\rho_c$ exhibits two peaks of opposite signs
at around $T_1$ and $T_2$. The change in sign suggests significant
modification in transport mechanisms at different temperatures. The
hysteresis observed in resistivity is an indication of first order
phase transition. Notably, the transition from paramagnetic to
C-type antiferromagnetic phase\cite{ch} is also observed at $T_1$.

The heat capacity data shown in Fig. 1(c) exhibit a sharp peak at
$T_1$. The magnetic entropy change at this transition obtained by
subtracting the polynomial background is found to be very large, 4.3
J/mol.K. No distinct feature is observed around $T_2$.
Interestingly, at low temperatures, $C_p/T$ exhibit a linear
dependence with $T^2$ as shown in Fig. 1(d). Considering $C_p/T =
\gamma +\beta T^2$, the value of $\gamma$ is found to be 5.6
mJ/mol.K$^{2}$. In metals, $\gamma$ corresponds to the electronic
contribution to $C_p$. However, the compound in this study exhibits
insulating temperature dependence. Thus, the observation of large
$\gamma$ indicating finite density of states at the Fermi level is
unusual and needs further study to understand this effect.

In Fig. 2(a), we show the temperature evolution of the modulus of
FT[k$^{2}$$\chi$(k)]. The FT curves are uncorrected for the central
and back scattering phase shifts. The figure shows three prominent
peaks around 1.5, 2.9 and 3.4 \AA\ marked by A, B and C,
respectively. Peaks A and B correspond to Mn-O and Mn-Sr/La shells,
respectively, and peak C corresponds to Mn-Mn and multiple
scattering paths (Mn-O-Mn). While there is no apparent shift in the
peak position of the peaks B and C, their relative intensity
decreases with the decrease in temperature as shown by plotting the
maxima of B and C in Fig. 2(b). At $T >$ 270 K, the peak B is weaker
than peak C. For 270 K $> T >$ 230 K, the intensities are very
similar and at $T <$ 230 K, the intensity of peak B increases
significantly relative to the intensity of peak C. Peak A exhibits a
small shift towards the origin and an increasing trend in intensity
with the decrease in temperature. Since, $\chi(k)$ is a function of
bond length, $R_j$ and Debye-Waller factor, $\sigma_j$ for the
$j$-th coordination shell, such modification in FT amplitude
indicates significant changes in these two parameters with
temperatures. In order to obtain the detailed structural information
of these modifications, we have carried out the EXAFS fitting on
Mn-O, Mn-Sr/La and Mn-Mn coordination shells. Representative fit of
these shells are shown in Fig. 2(c) and 2(d). The typical R-factors
for the first and higher coordination shell fits are 0.003 and
0.015, respectively.

First, we start with the first coordination shell corresponding to
Mn-O bonds. This shell is fitted with 2 short bonds, $R_1$ and 4
long bonds, $R_2$ and the results are shown in Fig. 3(a). It is
evident from the figure that the distribution in Mn-O bonds persists
with the decrease in temperature until 270 K. Interestingly, around
the region of phase transition ($\sim$ 265 K), R$_{1}$ and R$_{2}$
are almost identical indicating an undistorted MnO$_{6}$ octahedron.
At low temperatures, $R_1$ becomes significantly smaller than $R_2$
thereby MnO$_{6}$ octahedra become distorted. This is remarkable and
reveals presumably for the first time such behavior as a function of
temperature across a first order phase transition. This suggests
that tetragonal distortion starts from a perfectly cubic structure,
where the MnO$_6$ octahedra are also symmetric. A magnetic phase
transition (from PM to C-AFM) also occurs at $T_1$. Interestingly,
in sharp contrast to the common belief, the Mn-atoms corresponding
to the shorter bond length, $R_1$ becomes ferromagnetically coupled
and the ones with the longer bond lengths, $R_2$ are
antiferromagnetically coupled.

When the MnO$_6$ octahedra are symmetric, the $d_{xy}$, $d_{yz}$ and
$d_{xz}$ orbitals are degenerate forming a $t_{2g}$ band, and
$d_{x^2-y^2}$ and $d_{z^2}$ orbitals are degenerate forming an $e_g$
band. In the ionic configuration, Mn 3$d$ bands in
La$_{0.2}$Sr$_{0.8}$MnO$_3$ have 3.2 electrons and the up spin
$t_{2g}$ band will be completely filled. The structural changes in
the MnO$_6$ octahedra will lead to a change in the degeneracy of the
$t_{2g}$ and $e_g$ bands, and hence significant fluctuation in
occupancy of $d_{x^2-y^2}$ and $d_{z^2}$ orbitals. Since, the
octahedra are compressed along apical direction ($R_1$; say
$z$-direction), the $d$ orbital corresponding to the plane having
larger Mn-O bond lengths ($R_2$; $d_{x^2-y^2}$) will be more
populated compared to the perpendicular one. This is presumably the
reason for the anomalous magnetic coupling observed in this system
as filled up spin $d_{x^2-y^2}$ band favors antiferromagnetic
coupling due to superexchange interaction.

In order to explore plethora of unusual effects such as the sharp
peak in $C_p$, a steep increase in resistivity and the nucleation of
unusual magnetic coupling at $T_1$ further, we show the Debye-Waller
factors corresponding to $R_1$ and $R_2$ in Fig. 3(d) and 3(f). The
Debye-Waller factor corresponding to $R_1$ shown in Fig. 3(d) is
significantly large. This suggests strong influence of disorder in
localizing the Mn-O hybridized electronic states corresponding to
$R_1$. This is presumably the reason for insulating transport
despite the fact that ferromagnetic coupling along $R_1$ often
favors metallic conduction and finite $\gamma$ corresponds to finite
density of states at the Fermi level.

In Figs. 3(b) and 3(c), we show the Mn-Sr/La and Mn-Mn bond lengths,
respectively and corresponding Debye-Waller factors are shown in
Fig. 3(e) and 3(f), respectively. It is evident from the figures
that around T$_{1}$, the average Mn-Sr/La and Mn-Mn bond lengths
remain almost the same.

At 245 K, MnO$_{6}$ octahedra are the most distorted ones and the
Mn-Mn distance also becomes minimum. Below this temperature, the
distortion in MnO$_6$ octahedra reduces gradually with the decrease
in temperature and subsequently, the Mn-Mn distance also increases
gradually. The Debye-Waller factor corresponding Mn-Mn distance does
not change significantly (see Fig 3(f)) and that corresponding to
Sr/La atoms around Mn shown in Fig. 3(e) exhibits a gradual decrease
with temperature. This indicates that the disorder in Sr-sublattice
may not have significant role in the electronic properties of this
compound.

Around T$_{1}$, the value of resistivity during warming cycle
($\rho$$_{w}$) is more than the resistivity during cooling cycle
($\rho$$_{c}$). Below 257 K, the value of $\rho$$_{c}$ becomes more
than $\rho$$_{w}$ and a change in slope is also observed. Such
unusual modification in transport behavior presumably due to the
temperature induced structural evolution of the MnO$_6$ octahedra
leading to fluctuation of the occupancy of the Mn $e_g$ orbitals in
the presence of strong disorder.


In summary, we have studied the electronic properties of
La$_{0.2}$Sr$_{0.8}$MnO$_{3}$ using resistivity and heat capacity
measurements. Transport and heat capacity data reveal a first order
phase transition around 265 K. The local structure was probed by Mn
K-edge EXAFS technique. The MnO$_6$ octahedra are found to be
distorted even in the cubic phase at room temperature. EXAFS results
at different temperatures manifests remarkable evolution of the
MnO$_6$ octahedra across the first order phase transition. We
observe that the tetragonal distortion and the antiferromagnetic
ordering nucleates from a perfectly cubic structure, where the
MnO$_6$ octahedra are symmetric. Thus, the phase transition in this
compound involves significant fluctuation in the Mn 3$d$ orbital
occupancy and disorder induced localization, which may be important
in understanding interesting orbital and spin ordering observed in
manganites.

\pagebreak

\pagebreak

\section{Figure Captions:}

Fig. 1: (color online) (a) The resistivity during warming (open
circles), cooling (line) cycles. (b) The difference in the
resistivity during the cooling ($\rho$$_{c}$) and warming
($\rho$$_{w}$) cycles normalized by $\rho_c$. (c) The heat capacity,
$C_p$ as a function of temperature. (d) $C_p/T$ vs $T^2$ ($T$ =
temperature) shown in the low temperature region.

Fig. 2: (color online) (a) The Fourier transform (FT) of $k^2
\chi(k)$ at different temperatures. (b) The intensities of the peaks
A, B and C as a function of temperature. Representative fits for the
case of (c) first and (d) higher coordination shells at 255 K. The
open circles and the solid line indicate the experimental and the
fitted patterns respectively.

Fig. 3: (color online) The variation of (a) $R_1$ and $R_2$, (b)
Mn-Sr/La and (c) Mn-Mn bond lengths as a function of temperature
obtained from EXAFS fitting. The Debye-Waller factors corresponding
to (d) $R_1$, (e) Sr/La, and (f) $R_2$ and Mn as a function of
temperature.

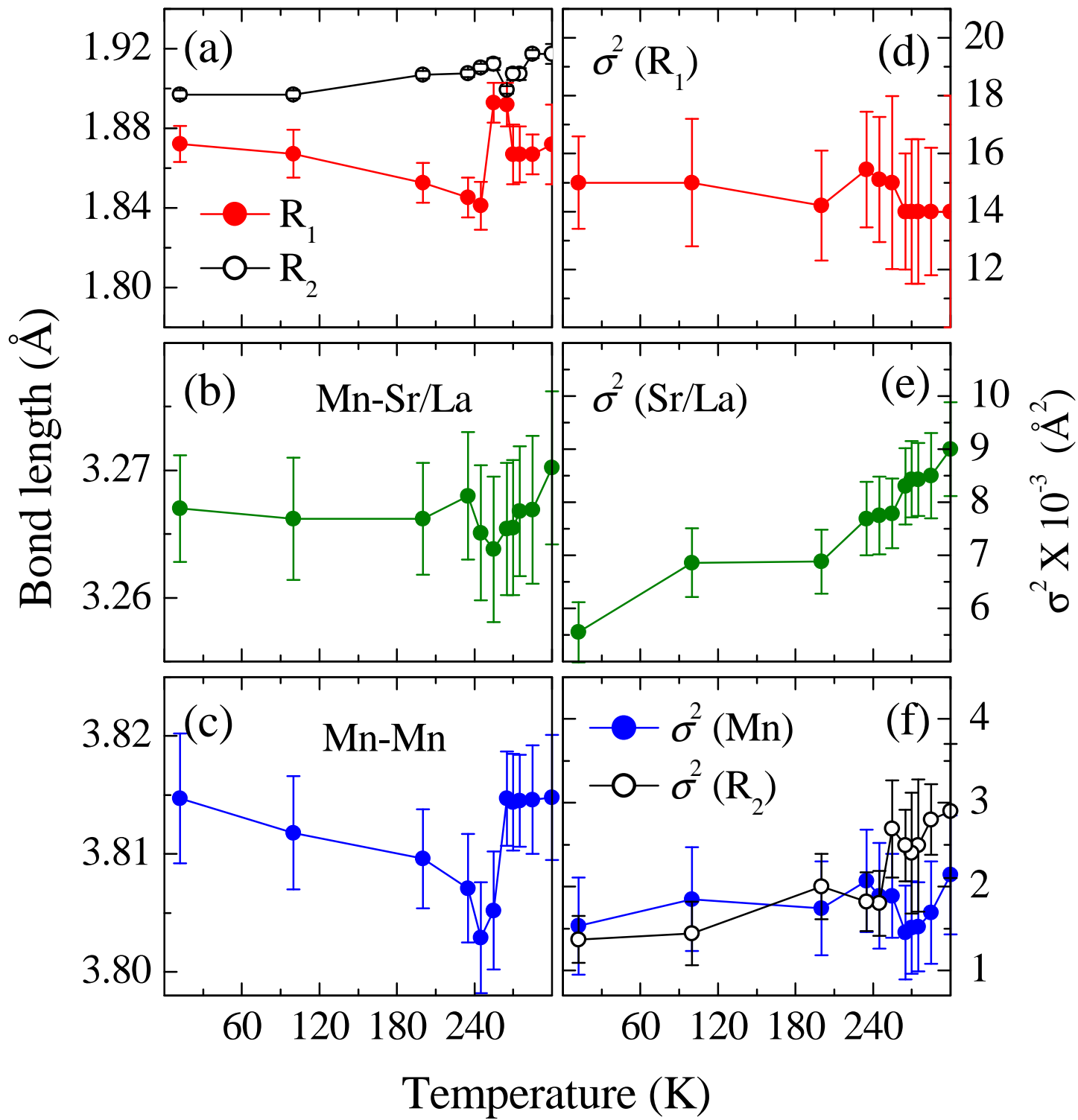
\end{document}